\documentclass{article}

\usepackage[table]{xcolor}
\usepackage{soul}
\usepackage{titlesec}
\usepackage{setspace}
\usepackage{authblk}
\usepackage{url}
\usepackage{tikz}
\usepackage{tikzscale}
\usepackage{hyperref} 
\usepackage{pgfplots}
\usepackage{caption}
\usepackage{algorithm}
\usepackage{algorithmic}
\usepackage{amsmath}
\usepackage{amssymb}
\usepackage{tabularx}
\usepackage{tabulary}
\usepackage{multirow}
\usepackage{scalefnt}
\pgfplotsset{width=10cm,compat=1.15}
\usepackage{enumitem}
\usepackage{bbm}

\setlist{nolistsep}

\usepackage{etoolbox}

\titlespacing\section{0pt}{5 pt}{5pt}
\titlespacing\subsection{0pt}{5 pt}{5pt}
\titlespacing\subsubsection{0pt}{5 pt}{5pt}

\usepackage{amsmath}
\usepackage{amsthm}
\usepackage{amssymb}
\usepackage{amsfonts}
\usepackage{amsxtra}     
\usepackage{verbatim}
\usepackage{mathrsfs}
\usepackage{accents}

\usepackage{algorithm}
\usepackage{algorithmic}

\makeatletter
\newcommand{\ALOOP}[1]{\ALC@it\algorithmicloop\ #1%
	\begin{ALC@loop}}
	\newcommand{\ENDALOOP}{\end{ALC@loop}\ALC@it\algorithmicendloop}

\makeatother

\usepackage{nomencl}
\makenomenclature
\usepackage{etoolbox}
\renewcommand\nomgroup[1]{%
  \item[\itshape\bfseries
  \ifstrequal{#1}{N}{Set}{%
  \ifstrequal{#1}{P}{Parameters}{%
  \ifstrequal{#1}{D}{Decision variables}{
  \ifstrequal{#1}{V}{Variable}{}{}}}}%
]}

\allowdisplaybreaks

\usepackage[comma]{natbib}
\setcitestyle{authoryear}
\title{Beyond capacity: contractual form in\\ electricity reliability obligations}

\author[1]{Han Shu\thanks{hs2226@cornell.edu}}
\author[2]{Jacob Mays\thanks{jacobmays@cornell.edu}}

\affil[1]{Systems Engineering, Cornell University, Ithaca, NY 14853}
\affil[2]{Civil and Environmental Engineering, Cornell University, Ithaca, NY 14853}

\begin{document}

\maketitle

\begin{abstract}
Liberalized electricity markets often include resource adequacy mechanisms that require consumers to contract with generation resources well in advance of real-time operations. While administratively defined mechanisms have most commonly taken the form of a capacity obligation, efficient markets would feature a broad array of arrangements adapted to the risk profiles and appetites of market participants. This article considers how the financial hedge embedded in alternative resource adequacy contract designs can induce different responses from risk-averse investors, with consequences for the resource mix and market structure. We construct a stochastic equilibrium model describing a competitive market with incomplete risk trading and compute investment equilibria under different contracting regimes. Two policy recommendations result. First, to avoid creating inefficiency by crowding out other forms of risk sharing, system operators should allow resources contracted through other means to opt out of mandatory capacity mechanisms, with their contribution to those requirements subtracted from administratively defined demand curves. Second, if they wish to promote a single contractual form, regulators should consider replacing existing option-like capacity mechanisms with a shaped forward contract for energy. Beyond these recommendations, we discuss the tension that liberalized systems face in seeking to promote both reliability and competitive outcomes.
\end{abstract}

\textbf{Keywords:} Electricity market design, capacity markets, risk aversion, cost of capital, resource adequacy
\onehalfspacing

\section{Introduction}
Electricity spot prices are so volatile as to threaten the viability of markets constructed on their basis. In February 2021, Winter Storm Uri led to unprecedented supply shortfalls in the Electric Reliability Council of Texas (ERCOT), leading many observers to question the role of market design in contributing to the system's lack of resilience~\citep{baker2022paying,Gruber2022,mays2022private}. In June 2022, the Australian National Electricity Market (NEM) was suspended for a ten-day period as the Australian Energy Market Operator (AEMO) claimed that it had become impossible to secure electricity through the market~\citep{AEMO}. On a longer timescale, high gas prices and supply interruptions due to the ongoing Russo-Ukrainian War have to led to an energy crisis in Europe, with political leadership in several countries calling for a return to the average cost pricing characteristic of regulated monopolies rather than marginal cost pricing inherent to competitive markets~\citep{EUR_COMMISSION}.\par 

Both for consumer protection and to avoid the underinvestment that can result from financial and political instability, regulators and system operators have an interest in mitigating the effects of volatility. In the first instance, it might be questioned why there must be any centrally designed component to risk sharing, as opposed to counting on market participants to allocate risk efficiently according to their preferences~\citep{hogan2005energy}. The first answer to this question is based on ``missing money'' and the second on ``missing markets.'' First, due to operator interventions and market power mitigation measures, many jurisdictions fail to produce spot prices that are high enough to support an adequate level of resources~\citep{joskow2008capacity,Cramton2013}. Lacking full-strength spot prices, generators need a supplemental revenue stream to support investment. Most commonly, this supplemental revenue takes the form of a capacity payment. While design details vary substantially across jurisdictions, the financial effect of capacity payments is similar to that of a call option: generators receive a consistent monthly or annual payment in exchange for the potential for the high prices they would otherwise see in times of scarcity. The particular form of risk transfer implied by capacity payments is a side effect of the chosen response to the missing money problem, rather than a deliberate design choice. Even if spot prices are high enough to avoid the missing money problem, however, long-standing concern about missing markets for risk sharing means that some form of obligation that retailers contract with generation resources could be warranted ~\citep{Neuhoff2004,Newbery2016,mays2022private}. In both Texas and Australia, the two most prominent examples of markets that have committed to full-strength scarcity prices and avoided contracting obligations to date, recent events have motivated regulators to reexamine some form of capacity mechanism. \par

In this context, the goal of this paper is to consider the effects that different forms of contracting obligations may have on the resource mix, market structure, cost, and reliability outcomes in liberalized systems. In theory, achieving near-optimal investment outcomes requires a menu of contracts adapted to the risk profiles of different market participants~\citep{willems2010market}. In this vein, there is no theoretical reason to privilege the particular option-like contractual form embedded in capacity markets. Because the payout structure built into capacity markets is well-adapted to peaking plants,~\cite{mays2019asymmetric} argues that existing mechanisms can tilt the resource mix in that direction. The European crisis highlights that although they can improve resource adequacy, capacity mechanisms offer relatively little price protection to consumers and represent an incomplete solution to the problem of missing markets~\citep{batlle2022b,batlle2022a}.  More concretely, we address two questions of immediate relevance to resource adequacy debates. First, if capacity markets continue in something like their current form, how should variable and energy-limited resources contracted through other means participate in them? Second, would the standardized fixed price forward contract proposed in~\cite{wolak2022long} offer a superior alternative to current capacity markets? \par

To address these questions, we construct a stochastic equilibrium model describing a competitive market with incomplete risk trading and compute investment equilibria under different contracting regimes. In the model, generation investors build capacity and take long-term contract positions in order to maximize risk-adjusted profits across potential future operating scenarios. Mandatory contracts condition the residual risk that must be managed by market participants themselves through complementary instruments. Accordingly, the equilibria resulting depend not only on the mandated contract but also on assumptions about hedging by other means. Under the extreme assumption of complete markets in risk, an obligation imposed by the regulator or system operator may have no effect: participants could simply unwind their mandated position through other channels. Without this assumption, the analysis requires characterizing the potential responses of participants to their residual risk exposure.\par

The appropriate treatment of variable renewables in capacity markets has attracted substantial recent attention along two lines. The first line of work measures how much capacity non-traditional resources should be allowed to sell, e.g., on the basis of their effective load carrying capability~\citep{bothwell2017crediting,schlag2020capacity}. From an idealized economic perspective, the need to assess a capacity value emerges from failures in the design of capacity markets: if penalties in the capacity market were strong enough, then any resource should only be willing to take on an obligation that they are able to deliver on. With risk neutral investors, efficiently calibrated non-performance penalties for capacity would provide the same incentive as full-strength spot prices for energy~\citep{Cramton2013}. In practice, while many markets in the U.S. have recently moved to strengthen non-performance penalties, they still fall short of what is required in theory. Throughout the numerical studies we exclude the possibility of bankruptcy and retain full-strength spot prices, providing an efficient penalty for non-performance on contracts~\citep{Vazquez2002,oren2005generation,hogan2005energy}, but we return to the implications of weaker penalties later in the discussion. The second line is how to incorporate resources supported by state-initiated contracts~\citep{Macey2021}. Seen in the context of missing markets for long-term risk sharing, these contracts have typically supported low-carbon technologies that are disadvantaged by capacity markets. Accordingly, while in the U.S. some have argued that these contracts interfere with price formation in capacity markets, our modeling framework treats them as complementary instruments leading to a more complete ``hybrid'' market~\citep{Joskow2022}. This more optimistic view, however, depends on an assumption that the state-initiated contracts can be treated not merely as subsidies, but instead as hedges that are efficiently incorporated in the portfolios of risk-averse load-serving entities~\cite{simshauser2019stability}.\par

Our results suggest that instead of current attempts to fit variable resources into capacity markets, it is more efficient to simply remove them and correspondingly reduce the administratively defined demand for capacity. In a risk neutral analysis, the two approaches are identical. When risk aversion is considered, however, compelling a variable resource to participate in the capacity market necessitates that they sell a contract that would be unlikely to arise in any self-organizing market. Both in theory and in practice, it is unnatural to expect variable resources to sell hedging instruments with a constant volume. When securing contracts, wind and solar producers typically seek to match the shape of their offtake agreements with the shape of their production, e.g., through a proxy generation power purchase agreement. In taking on a capacity obligation, by contrast, wind and solar make a commitment they can only physically support some fraction of the time. With theoretically efficient non-performance penalties, selling capacity leaves variable generators exposed to significant non-performance risk. Accordingly, attempting to fit the square peg of variable generators in the round hole of capacity markets can create inefficiencies, with generators charging risk premia or withholding supply from the market. A related prediction is that the willingness of variable resources to participate in current capacity markets is in part a reflection of the relatively weak penalties they employ.

While the first question implies a more moderate reform that keeps existing capacity markets in place, the second envisions a more significant overhaul of resource adequacy toward long-term contracts for energy, along the lines of those used in Chile and Brazil~\citep{Moreno2010,Munoz2021}. Motivated by the context of California,~\cite{wolak2022long} describes a standardized fixed-price forward contract (SFPFC) approach in which suppliers sell a quantity of energy at a given price, with the shape of the contract determined ex post to match the actual consumption of loads on the system over the course of the delivery period. In this way, the contract mimics a full requirements contract but covers a fixed volume, leaving incentives on the margin intact. The proposal envisions auctions for these standardized contracts occurring on a regular basis, with retailers required to cover an increasing percentage of its expected load as the relevant delivery period approaches. A key difference from capacity markets is that the price of energy delivered under the contract is determined in advance; while capacity obligations require suppliers to offer in real time, they typically do not specify a price at which they must offer. In the SFPFC approach both price and shape risk are shifted from retailers to suppliers, resulting in a much stronger hedge for consumers. For this reason, the SFPFC may be particularly attractive in jurisdictions without retail competition.\par

Compared against the option contract we use to mimic a capacity market, SFPFCs result in equilibria with greater surplus, lower cost for comparable reliability, and substantially lower volatility in the price seen by consumers. In other words, if regulators and system operators wish to promote a single instrument for long-term risk sharing, the SFPFC approach may be seen as superior to capacity markets. However, two caveats apply. The first is that, while SFPFCs outperform options when each is the only contract available, they are not able to match the efficiency achievable with a larger number of instruments. The second is that, since suppliers of SFPFCs bear risk related to both price and shape, there is an incentive for suppliers to create portfolios of assets protecting against both. As such, just as missing markets for risk sharing between generators and retailers can lead to vertical integration in energy-only markets~\citep{Simshauser2021}, the SFPFC design could create pressure toward more consolidation among suppliers. 

While the mandatory contracts considered here can improve resource adequacy, they cannot completely resolve reliability risks along the electricity value chain. First, our analysis neglects distinctions between reliability needs at the customer and device level (see, e.g.,~\cite{Billimoria2019,Billimoria2022}). Second, a point of emphasis in the context of the 2022 European energy crisis as well as the 2021 Texas outages is the security of supply of input fuels, most importantly natural gas. Mandatory contracting between generators and retailers for resource adequacy implicitly assigns the responsibility for ensuring security of supply to the generators, who will be unable to deliver on their obligations if they fail to procure input fuel. While this assignment of responsibility may be an improvement over a situation in which no entity has clear responsibility, it does not in itself resolve the issue. With that said, programs for resource adequacy are clearly implicated in security of supply to the extent that they encourage dependence on different fuels. Third, while our analysis of the first question posed above emphasizes non-performance risk for variable resources, traditional thermal resources are also subject to unexpected failures. Unless generators of all types can efficiently insure against non-performance, there is a potential for excessive risk premia or the use of bankruptcy as a hedge. \par

\section{Mandatory Contracting and Portfolio Crowding}

The results center on two sets of numerical examples employing a two-stage stochastic equilibrium model describing capacity investment in a perfectly competitive market with risk-averse generation investors. In the first stage, agents make investments and financial trading decisions. In the second stage, system operators perform a year-long economic dispatch to determine spot prices and generation output. Details of the equilibrium modeling framework are given in the Methods. Compared to previous models of this form~\citep{ehrenmann2011generation,Ralph2015,Philpott2016,DAertrycke2017wp,Abada2017,mays2019asymmetric}, our most important change arises in our second set of examples, in which we allow investors to build portfolios containing multiple technologies rather than having each technology represented by a single representative agent. This change enables a discussion of potential changes to market structure resulting from contracting mandates.\par

Our first set of numerical examples considers the interaction between mandatory capacity markets and other forms of risk trading, such as government-initiated contracts for differences, corporate power purchase agreements, and bank hedges. As in \cite{mays2019asymmetric}, the stylized system has three resources available: baseload, with a high upfront cost but low and certain operating cost, peaking, with a low upfront cost but high and uncertain operating cost, and variable, with a low upfront cost and zero operating cost but uncertain availability. Whereas \cite{mays2019asymmetric} considers the effect of capacity markets in the absence of complementary forms of risk trading, here we assume that complementary trades are available but participation in the capacity market is nevertheless mandated. Imposition of a mandatory contract changes the risk profile of market participants, affecting the demand for other physical and financial assets. In our setting, the additional constraint degrades surplus and pushes the equilibrium capacity mix away from the variable resource.\par

For the numerical experiments, the nominal demand is based on the load curve of PJM for one year with hourly resolution. The three sources of uncertainty are fuel cost, demand, and the availability of the variable generator. Fuel costs and demand shifts are modelled as random variables with 10 scenarios each, while the availability of variable generation is modeled with 4 scenarios, leading to a total of 400 scenarios for the second-stage economic dispatch with equal probability. The four availability profiles are generated with an average availability of 37.5\% and vary in their correlation with the fixed load. To characterize the risk attitudes of market participants we use a weighted sum of the expected value of surplus, with weight~$\beta$, and the conditional value at risk (CVaR) of the $\alpha$ tail of the surplus, with weight $1-\beta$. This weighted sum gives a coherent risk measure~\citep{Artzner1999}, convenient for implementation in an optimization model~\citep{Rockafellar2000}. The risk parameters are set as $\alpha=\beta=0.7$ for retailers and $\alpha=0.7$, $\beta=0.2,0.4,0.6,0.8$ for generators, with higher values of $\beta$ implying less risk aversion. 

With complete markets in risk, participants would be able to hedge though a set of 400 Arrow\textendash Debreu securities corresponding to the 400 second-stage scenarios. With this assumption, a competitive equilibrium can be found by solving a risk-averse optimization problem\citep{ehrenmann2011generation,Ralph2015,Philpott2016,Ferris2022}. The complete trading model and the resulting equilibrium is described in Supplementary Note 1. With three sources of uncertainty, the system is able to achieve near-optimal results with three contracts that are adapted to risk profiles of the three generation technologies: a call option with strike price of \$1000/MWh, a future settling at \$50/MWh, and a unit contingent contract that matches the availability of the variable generator. Throughout the paper, we refer to this three-contract configuration as ``unrestricted" trading and treat it as the benchmark against which surplus in other configurations is measured. All surplus comparisons are calculated within the same risk parameters; i.e., while greater risk aversion leads to lower surplus in expectation, our focus is not on this effect but rather on the influence of different contracting regimes. In the unrestricted case, retailers are responsible for signing different types of contracts with generators to assemble a near-efficient portfolio.\par


As a contrast with the unrestricted case, we construct a ``mandatory'' case in which retailers are obligated to purchase a certain minimum number of option contracts. The minimum quantity is set based on the effective load carrying capability of the equilibrium mix found in the unrestricted case, implying roughly equivalent reliability in the two settings. Two additional changes are needed to mimic a capacity market. First, we redefine the future and unit contingent contracts to pay only up to the option strike price, so that a generator does not ``double sell" the revenue associated with scarcity prices. In this way, generators sell their output in two tranches: an energy component up to option strike price and a capacity component above the strike price. This is analogous to capacity markets where generators receive capacity payments and suppressed energy payments. Second, to avoid generators simply selling options in a speculative way to satisfy the purchase requirement, we also impose a limit on the maximum quantity of options that generators can sell. This capacity value is calculated for each resource based on its average availability during scarcity, reflecting the contribution of different generation resources to reliability.\par

Table~\ref{table:unrestricted_vs_mandatory} shows the impact of introducing the mandatory option on the resource mix and prices arising in the system. At all levels of $\beta$, the resource mix shifts away from the variable and peaking and toward the baseload resource, with overall reliability in the system roughly unchanged. A consequence of this changed mix is degradation in surplus, ranging from \$1,044M with $\beta=0.2$ to \$389M with $\beta=0.8$. As the generators become less risk averse (i.e., with higher values of $\beta$), the effects of risk aversion and the contract mandate become smaller, with the resource mix in the two cases converging and the loss of surplus shrinking. The consumption-weighted average price paid by consumers including contracts is slightly higher than spot prices, reflecting the net effect of risk premia on the three contracts. Under both contracting regimes, consumers see relatively little of the underlying volatility in spot prices.

\begin{table}
\captionsetup{justification=raggedright,singlelinecheck=false,labelfont=bf}
\begin{center}
\caption{\textbf{Effect of mandatory options contract.} Four pairs of equilibria are shown in order of decreasing risk aversion (higher $\beta$). For each $\beta$, introducing a minimum quantity for the options contract degrades surplus and pushes the equilibrium mix away from the variable resource. Volatility is calculated as the standard deviation in consumption-weighted prices in the second-stage operating years. Change in surplus is calculated against the unrestricted case at the same level of risk aversion.}\label{table:unrestricted_vs_mandatory}
\begin{tabular}{l|ccccccc}
\hline 
\hline
& \multicolumn{2}{c}{$\beta=0.2$} & \multicolumn{2}{c}{$\beta=0.4$}  \\
 & Unrestricted & Mandatory & Unrestricted & Mandatory \\
\hline
Capacity (GW)   \\
\quad Baseload &  31.1  & 58.3  & 32.0  & 50.8 \\ 
\quad Peaker   &  107.4  & 88.9  & 106.8  & 93.6 \\ 
\quad Variable &  159.0  & 106.8  & 157.3  & 122.7 \\ 
\\
Average Price (\$/MWh) \\
\quad Spot  & 56.74 & 56.57 & 56.75 & 56.77 \\
\quad Hedged & 58.96 & 60.02 & 58.94 & 59.74 \\
\\
Interannual Volatility (\$/MWh) \\
\quad Spot & 23.26 & 23.22 & 23.23 & 23.61 \\
\quad Hedged & 3.27 & 3.87 & 3.23 & 4.17 \\
\\
Expected Unserved Energy (GWh) &  4.80  & 4.76  & 4.78  & 4.96 \\
Proximity to Equilibrium &  0.024\%  & 0.043\%  & 0.031 \% & 0.057\% \\
Change in Surplus (\$M/yr) &  --  & -1,044  & --   & -848 \\
\hline
 & \multicolumn{2}{c}{$\beta=0.6$} & \multicolumn{2}{c}{$\beta=0.8$}\\
 & Unrestricted & Mandatory & Unrestricted & Mandatory \\
\hline
Capacity (GW)   \\
\quad Baseload &  33.3   & 46.2  & 33.8   & 36.4 \\ 
\quad Peaker   &  105.7   & 96.6  & 105.1   & 103.7 \\ 
\quad Variable &  155.4   & 131.6  & 154.8  & 149.2 \\ 
\\
Average Price (\$/MWh) \\
\quad Spot  & 56.75 & 57.15 & 57.15 & 56.81 \\
\quad Hedged  & 58.93 & 59.46 & 58.91 & 59.34 \\
\\
Interannual Volatility (\$/MWh) \\
\quad Spot & 23.33 & 24.12 & 24.01 & 23.19 \\
\quad Hedged & 3.15 & 4.41 & 3.07 & 2.38 \\
\\
Expected unserved energy (GWh) &  4.85   & 5.27  & 5.25  & 4.76 \\
Proximity to Equilibrium &  0.15\%  & 0.098\%  & 0.049\%  & 0.054\% \\
Change in Surplus (\$M/yr) &  --   & -662 & --   & -389\\
\hline
\hline
\end{tabular}
\end{center}
\end{table}

Contract volumes and risk premia are reported in Table~\ref{table:unrestricted_trading}. In the unrestricted case, each technology primarily trades contracts adapted to its risk profile: baseload plants sell futures, peaking plants sell options, and variable plants sell unit contingent. Peakers sell options beyond their physical capacity, indicating a purely financial aspect to the trade. In the mandatory option case, baseload is able to fully hedge its production by pairing options with an equal amount of the redefined futures. When $\beta =0.8$, baseload sells futures slightly beyond what it is able to produce, with the willingness to engage in the speculative trade reflecting lower risk aversion. The peaker sells only options in most settings, adding a small number of futures only at higher levels of $\beta$. Similarly, when $\beta =0.8$ the variable resource sells futures instead of unit contingent contracts in the mandatory case, indicating a willingness to take a position that is less well matched to its physical capability in exchange for higher risk-adjusted profits. Because consumers do not absorb the shape risk of the variable generator, this configuration corresponds to the lowest standard deviation of prices paid among the eight shown in Table~\ref{table:unrestricted_vs_mandatory}. While risk premia for the futures and unit contingent contracts are not directly comparable due to the redefinition of both to exclude scarcity rents, the options contract is unchanged. At every level of $\beta$, the risk premium is higher under the mandatory case, reflecting the higher demand for this contract resulting from the obligation placed on retailers.

\begin{table}
\captionsetup{justification=raggedright,singlelinecheck=false,labelfont=bf}
\begin{center}
\caption{\textbf{Contract volumes and risk premia in three-contract cases.} Generators primarily trade the contracts adapted to their risk profile. When options are mandated, baseload resources are able to fully hedge its production but variable resources are not. Risk premia for the options contract grow due to the purchasing obligation.}\label{table:unrestricted_trading}
\begin{tabular}{l|ccccccc}
\hline 
\hline
& \multicolumn{2}{c}{$\beta=0.2$} & \multicolumn{2}{c}{$\beta=0.4$}  \\
 & Unrestricted & Mandatory & Unrestricted & Mandatory \\
\hline
Trade Volume (GW) \\

\quad Baseload \\
\qquad Futures &  27.9  & 52.5  & 28.8  & 45.7 \\ 
\qquad Options &  2.5   & 52.5  & 2.6  & 45.7 \\ 
\qquad Unit Contingent   &  0  &  0 & 0  & 0  \\ 
\quad Peaker \\
\qquad Futures &  0  &  0 & 0  & 0  \\ 
\qquad Options &  120.0   & 80.0  & 119.4  & 84.3 \\ 
\qquad Unit Contingent &  0  &  0 & 0  & 0   \\ 
\quad Variable \\
\qquad Futures &  0   & 0  & 0   & 0  \\ 
\qquad Options &  0   & 16.0  & 0  & 18.5  \\ 
\qquad Unit Contingent  &  159.0 & 106.8  & 157.3   & 122.7   \\ 
\\
Contract Risk Premium (\$/MW-yr) \\
\quad Futures & 16,470 & 5,638 & 16,113 & 5,839 \\
\quad Options & 6,747 & 17,366 & 6,738 & 14,597 \\
\quad Unit Contingent & 3,455 & 693 & 3,315 & 773 \\
\hline
 & \multicolumn{2}{c}{$\beta=0.6$} & \multicolumn{2}{c}{$\beta=0.8$}\\
 & Unrestricted & Mandatory & Unrestricted & Mandatory \\
\hline
Trade Volume (GW) \\
\quad Baseload \\
\qquad Futures &  30.0  & 41.6  & 31.2  & 33.7  \\ 
\qquad Options &  3.4   & 41.6  & 4.1  & 32.8 \\ 
\qquad Unit Contingent   &  0  &  0 & 0  & 0  \\ 
\quad Peaker \\
\qquad Futures &  0  & 0.08  & 0 & 0.3 \\ 
\qquad Options &  119.5  & 86.9  & 121.4  & 93.4 \\ 
\qquad Unit Contingent   &  0  &  0 & 0  & 0  \\ 
\quad Variable \\
\qquad Futures &  0   & 0  & 0  & 59.6 \\ 
\qquad Options &  0   & 20.1  & 0  & 22.4 \\ 
\qquad Unit Contingent   &  155.4   & 131.6  & 154.8 & 0.0002 \\ 
\\
Contract Risk Premium (\$/MW-yr) \\
\quad Futures & 16,033 & 6,646 & 13,481 & 5,524 \\
\quad Options & 6,261 & 10,108 & 3,961 & 10,870 \\
\quad Unit Contingent & 3,451 & 1,171 & 3,185 & 1,125 \\
\hline
\hline
\end{tabular}
\end{center}
\end{table}

\section{Load-shaped Contracts and Portfolio Optimization}

While the examples in the previous section compare the contract mandate against a situation of ``near-complete'' markets, the larger concern in many cases is market incompleteness. A greater degree of incompleteness expands the scope for regulatory interventions to improve outcomes. Perhaps the least interventionist approach currently in use is Australia's Retailer Reliability Obligation (RRO), which allows the regulator to trigger a requirement that retailers hold contracts sufficient to serving their load, without specifying the particular form of those contracts. The RRO had not been triggered before the NEM's recent suspension, indicating that its design was insufficiently interventionist. An enhanced version of the RRO could simply be active at all times rather than requiring a trigger. In this case the system would still rely on decentralized contracting, subject to a centralized determination of the quality of different contracts and the total that must be held. Most systems instead define a particular contractual form (e.g., a capacity product), potentially promoting liquidity and market depth at the expense of flexibility.\par 
The examples in this section assume that market participants share risk only through a single, centrally defined contract and investigates the question of what contractual form to choose. A targeted approach to defining a single product would be to identify the dimension along which incompleteness is the most salient and design an instrument along that dimension, leaving market participants to manage residual risk with supporting contracts. A more holistic approach, as in the SFPFC described in~\cite{wolak2022long}, is to attempt maximal coverage of risk for consumers. The examples in this section compare the SFPFC with the option contract we use as a proxy for conventional capacity markets. In the SFPFC, the total quantity of covered energy is determined in advance for a delivery period (in our case, one year), but its shape is determined ex post to match consumer demand. Compared with call options, this approach provides a stronger hedge for retailers, covering a much larger portion of the expected cost to serve load. 

Whereas the ``unrestricted'' case in the previous section featured three contracts adapted to the three technologies in the system, the shape of the SFPFC is not well matched to the production of any of the three technologies. As a consequence, sellers have an incentive to construct portfolios containing a mix of the three technologies. To illustrate this incentive, we model two cases where generators can sell either separately or collectively. When selling SFPFC separately, each technology is left with shape risk given that none follows load precisely. By selling SFPFC collectively, the generation technologies are better positioned to deliver contracted quantities. In the numerical experiment, we treat the generation company as a single agent who owns all technologies and trades with retailers, modifying the solution approach to enable adjustments in the capacity mix held by this representative agent.\par

Table~\ref{tab:SFPFC} shows the equilibria resulting with the SFPFC sold separately and collectively, as well as a case where only options are traded. As should be expected, a single contract cannot match the performance of the three-contract ``unrestricted'' configuration shown in the previous section. However, the SFPFC clearly outperforms the option contract, especially when sold collectively, achieving social surplus that is much closer to the unrestricted trading case. This superior performance manifests primarily in lower prices; unserved energy is comparable between the options and collective SFPFC cases. The SFPFC has a particularly strong effect on interannual price volatility for consumers, bringing it below even the level of the three-contract case.\par

\begin{table}
\captionsetup{justification=raggedright,singlelinecheck=false,labelfont=bf}
\begin{center}
\caption{\textbf{Comparison of options against load-shaped forward contracts}. Change in surplus is calculate against the three-contract unrestricted case. SFPFC performs better than options when trading is limited to single instrument, particularly when sold collectively by a portfolio of technologies.} \label{tab:SFPFC}
\begin{tabular}{l|ccccccccc}
\hline 
\hline 
& \multicolumn{3}{c}{$\beta=0.2$} & \multicolumn{3}{c}{$\beta=0.4$}  \\
&Options & Sep & Col &Options & Sep & Col   \\
\hline
Capacity (GW)   \\
\quad Baseload &0.006 & 89.0 &40.7 & 0.02 &84.1 &34.6 \\ 
\quad Peaker   &137.3 & 72.5 &99.9 & 135.2 &75.0 &104.4   \\ 
\quad Variable &163.4 & 26.0 &143.2 & 176.2 &36.9 &153.6   \\ 
\\
Average Price (\$/MWh) \\
\quad Spot & 62.22 & 61.69 & 57.49 & 61.31 & 61.30 & 57.39\\
\quad Hedged &62.86 & 63.01 & 59.59 & 61.94 & 62.32 & 59.43  \\
\\
Interannual Volatility(\$/MWh) \\
\quad Spot & 25.86 & 22.33 & 24.89 & 25.42 & 23.41 & 24.43 \\
\quad Hedged & 13.79 & 1.02 & 1.75 & 13.63 & 1.17 &1.69\\
\\
Expected unserved energy (GWh) &5.54 & 4.05 &5.83 &5.53 &4.58 &5.52  \\
Proximity to Equilibrium &0.066\% & 0.058\% &0.016\% & 0.029\% &0.090\% &0.004\% \\
Change in Surplus (\$M/yr) &-5,042 & -3,500 &-517 &-4,206 & -2,906 &-375\\
\hline
&  \multicolumn{3}{c}{$\beta=0.6$} & \multicolumn{3}{c}{$\beta=0.8$}  \\
&Option & Sep & Col &Option & Sep & Col   \\
\hline
Capacity (GW)   \\
\quad Baseload &0.2 & 75.6 &33.6  &12.4 & 52.9 &31.4  \\ 
\quad Peaker  &133.6  & 79.6 &105.4 &122.6 & 92.8 &107.0  \\ 
\quad Variable &185.0 & 57.1 &154.3 &177.3 & 109.8 &157.7  \\ 
\\
Average Price (\$/MWh) \\
\quad Spot  & 60.41 & 61.00 & 57.44 & 59.40 & 60.23 & 57.59\\
\quad Hedged & 61.03 & 61.44 & 59.26 & 59.98 & 60.36 & 59.06 \\
\\
Interannual Volatility(\$/MWh) \\
\quad Spot & 25.18 & 24.38 & 24.20 & 24.63 & 26.85 & 24.34 \\
\quad Hedged & 13.61 & 1.28 & 1.38 & 13.45 & 1.36 & 1.14\\
\\
Expected unserved energy (GWh) &5.52 & 5.35 &5.35 &5.53 & 6.83 &5.39 \\
Proximity to Equilibrium &0.048\% & 0.071\% &0.008\% &0.073\% & 0.070\% &0.012\%\\
Change in Surplus (\$M/yr) &-3,411 & -2,183 &-212 &-2,448 & -1,243 &-41 \\
\hline
\hline 
\end{tabular}
\end{center}
\end{table}

Three aspects of the results in Table~\ref{tab:SFPFC} warrant further discussion. First, while the structure of the models constructed throughout this paper give rise to the possibility of multiple equilibria~\citep{Abada2017,Gerard2018}, we were able to identify multiple equilibria only in the case of SFPFCs sold collectively. Alternate equilibria have comparable surplus to those shown in Table~\ref{tab:SFPFC} and are discussed in Supplementary Note 2. Second, while the results show a large benefit when we allow SFPFCs to be sold collectively, no comparable benefit occurs when options can be sold collectively; results under this configuration are shown in Supplementary Note 3. Third, unserved energy is higher in the configuration with SFPFCs sold collectively than in the three-contract case. In Supplementary Note 4 we show results with a mandatory requirement on purchase of SFPFCs, enabling the system to achieve equivalent reliability with moderate loss of surplus.

Contract volumes and risk premia are shown in Table~\ref{tab:SFPFC_contract}. As in Table~\ref{table:unrestricted_trading}, risk premia tend to fall as the generators become less risk averse. Risk premia for the SFPFC are lower when sold separately, partially offsetting the effect of the higher underlying spot prices seen in this configuration. Trade volumes when selling SFPFC collectively are always smaller than when selling separately, indicating less hedging pressure when the technologies are combined in a portfolio. Whether sold separately or collectively, risk premia fall and trade volumes grow as the generators become less risk averse.

\begin{table}[]
\captionsetup{justification=raggedright,singlelinecheck=false,labelfont=bf}
    \centering
    \caption{\textbf{Contact volumes and risk premia in one-contract cases.} When technologies are allowed to sell SFPFC collectively, hedging pressure is lower and trade volumes decrease. Risk premia for the SFPFC are lower when sold separately, partially offsetting the effect of higher underlying spot prices.} \label{tab:SFPFC_contract}
    \begin{tabular}{l|l|ccccc}
\hline
\hline
& & $\beta=0.2$ & $\beta=0.4$ & $\beta=0.6$ & $\beta=0.8$  \\
\hline
\multirow{6}{*}{Options}&Trade Volume (GW) \\
&\quad Baseload &  0.0  &  0.0 & 0.2  & 12.5  \\ 
&\quad Peaker   &  153.0  &  150.7 & 149.0  & 137.0 \\ 
&\quad Variable &  7.4  &  8.2 & 10.1  & 8.2\\
& \\
&Contract risk premium (\$/MW-yr) &2,671 &2,643 &2,650 &2473 \\
\hline
\multirow{6}{*}{Sep}&Trade Volume (GW)  \\
&\quad Baseload  & 60.7 & 56.1 & 49.3 & 35.0  \\ 
&\quad Peaker   & 34.3 & 37.1 & 40.5 & 48.4 \\ 
&\quad Variable & 5.0 & 7.2 & 10.6 & 16.6 \\ 
& \\
&Contract risk premium (\$/MW-yr) & 9,057 & 6,564 & 1,782 & -582 \\
\hline
\multirow{3}{*}{Col}&Trade Volume (GW)  &91.6  &91.9 &93.6 &95.8 \\
& \\
&Contract risk premium (\$/MW-yr) &19,039  &18,335 &15,866  &12,256 \\
\hline
\hline
\end{tabular}
\end{table}

\section{Discussion} \label{se:conclusion}
The numerical results highlight that the manner in which a reliability obligations are defined can affect cost, reliability, and emissions outcomes in liberalized electricity markets. In the first set of experiments, imposing a capacity obligation crowds out other forward contracting and leads to worse outcomes for consumers. In the second, introducing a shaped forward contract instead of an option facilitates greater risk sharing and better outcomes for consumers. Taken together, the experiments demonstrate how the success of interventions facilitating or mandating risk sharing depends not just on the intervention itself but also on complementary mechanisms. Focusing on two policy recommendations, here we discuss the implications that these experiments have for debates about resource adequacy and consider aspects of the modeling that warrant further investigation.\par

The first recommendation concerns the incorporation of resources contracted through other means in mandatory capacity markets. In many jurisdictions, policymakers have introduced processes to award long-term contracts to renewables and storage. Beyond these state-initiated contracts, new resources are often able to contract directly with consumers of electricity (e.g., large firms with clean procurement goals). Such resources could be handled in either of two ways. First, the system operator could reduce the administratively defined demand for capacity, reflecting the lower probability of scarcity situations due to the presence of the resource. Second, the system operator could compel the resource to participate in the capacity construct. In risk-neutral terms the approaches should be equivalent, but when risk aversion is considered the first approach can be seen as preferable. It should be noted, however, that a change in approach would require changes in contractual terms: instead of receiving a payment for their capacity contribution directly, generators would need to monetize the decreased capacity obligation of offtakers, who may be seen as riskier counterparties than the system operator.\par

Rather than working around current capacity obligations, the second recommendation contemplates replacing them. Evaluated on their own, the modeling results suggest that SFPFCs promotes greater efficiency than current capacity obligations. However, several caveats apply. First, we note that the SFPFC as implemented in our model assumes the presence of full-strength spot prices and omits network constraints. As a result, we avoid two issues that have plagued real-world capacity markets, i.e., determination of non-performance penalties and resolving deliverability issues. Without ensuring full-strength spot prices, neither of these issues is resolved by a switch to the SFPFC. Second, as with the first recommendation, adopting this approach would create the need for changes in contractual arrangements that currently exist between generators and retailers. Since the SFPFC would hedge almost all consumption for retailers, a more natural counterparty for individual generators would instead be aggregators selling SFPFCs.

Beyond these recommendations, the results reflect a more fundamental challenge for resource adequacy in liberalized markets. The observation common to both sets of experiments is that generators will be reluctant to take on obligations that they are not able to defend physically. Since all of our examples use a single representative agent for each technology, non-performance risk is inherently pooled across the fleet. In reality, each individual unit would sell its own obligation, leading to greater risk for individual sellers of any resource adequacy product. Throughout the examples we retained full-strength scarcity prices and excluded the possibility of bankruptcy, ensuring efficient penalties for non-performance on contracts. However, the examples point to the challenges these full-strength prices or penalties present: greater pressure for consolidation among suppliers and a need for stronger credit requirements or an insurance mechanism (cf.~\cite{Billimoria2019}) to prevent the use of bankruptcy as a hedge. In this way, the relatively weak performance incentives attached to current capacity markets can be interpreted as a way to preempt problems with market power and creditworthiness. At the same time, weak prices and penalties imply inefficient direct incentives for operations and investment. These insufficient financial incentives in turn imply a need for active oversight of the physical system (e.g., through the capacity accreditation process) to ensure performance. In other words, a purely financial mechanism for mandatory contracting between generators and loads will not by itself resolve challenges with resource adequacy even if the political will to sustain full-strength scarcity prices is assumed.

Along these lines, the results point to the question of how to find an appropriate balance between financial incentives and direct oversight when trying to ensure reliability in competitive markets, and the related question of how this balance should evolve with a shift to carbon-free resources. By avoiding full-strength prices or penalties, market operators effectively pool some non-performance risk, easing bankruptcy pressures and reducing demands on contracting. Non-performance risk for wind and solar may be more straightforward to quantify than for traditional thermal resources, since their key failure mode (i.e., the weather) is easily observable by system planners and regulators. Moreover, since variable renewables can do little to control the sun or wind underlying their performance, systems reliant on these resources may have greater latitude for pooling risk on behalf of suppliers without altering their performance. However, in doing so they may retain an inherent anti-competitive bias against resources for which physical contributions are less easily audited. Aggregations of distributed energy resources, for example, could find it difficult to validate their contribution physically several years in advance of delivery, given that the precise resources composing the aggregation may not have been identified at that point. Such a concern is not merely theoretical, as seen in PJM's efforts to block the participation of the SOO Green HVDC Link in its capacity market. Since the line does not specify the generation resources behind the power it will deliver, PJM and its market monitor argue that they cannot count on performance from the new competitor at the level expected from incumbents. The unstated admission underneath this argument is that the financial performance incentives for SOO Green would be inadequate under current market rules. Given the disadvantages of interfering with full-strength prices, finding more effective ways to facilitate risk sharing is likely to remain a central question in market design for resource adequacy.



\section{Methods} \label{se:model}
\subsection{Modelling framework}

Largely following~\cite{mays2019asymmetric}, here we describe a model for risk-averse capacity equilibrium with incomplete financial markets. The numerical experiments consider three sources of uncertainty. We use the random variable $D^+$ with scenarios $s \in \mathcal{S}$ to represent a upward demand shift in all time blocks, $A_{gt}$ with scenarios $r \in \mathcal{R}$ to represent generation availability, and $C_{gf}^{\mathrm{EN}}$ and $D^-$ with scenarios $f \in \mathcal{F}$ to represent fuel cost and a downward demand shift that is positive correlated with fuel cost. This leads to of a total of $\lvert \mathcal{S} \rvert \times \lvert \mathcal{R} \rvert \times \lvert \mathcal{F} \rvert$ scenarios with equal probability. For each agent $a \in \mathcal{A}$, we calculate a risk measure $\rho_a$ as a convex combination of expected value and CVaR of surplus. In this approach, each agent performs a risk-averse optimization implicitly assigns higher probability to the events within its own $\alpha$-tail of outcomes. The models are implemented in Julia~\citep{Bezanson2017Julia} using JuMP.jl~\citep{DunningHuchetteLubin2017JuMP} and solved with Gurobi~\citep{gurobi}.

\subsection{Notation}
\doublespacing\noindent\textit{Sets:}
\begin{itemize}
\item[] $g \in \mathcal{G}$: set of all generation technologies
\item[] $t \in \mathcal{T}$: set of time blocks
\item[] $s \in \mathcal{S}$: set of demand shifter
\item[] $r \in \mathcal{R}$: set of generator availability profiles
\item[] $f \in \mathcal{F}$: set of scenarios for fuel prices
\item[] $a \in \mathcal{A}$: set of market participants (generators $g$ and consumers $c$)
\end{itemize} \bigskip
\noindent\textit{Parameters:}
\begin{itemize}
\item[] $B$: value of non-price-responsive load (\$/MWh)
\item[] $L_t$: length of time blocks $t$ (h)
\item[] $C_{g}^{\mathrm{INV}}$: amortized investment cost of generator $g$ per unit capacity (\$/MW)
\item[] $C_{gf}^{\mathrm{EN}}$: per unit production cost of generator $g$ under fuel cost scenario $f$ (\$/MWh)
\item[] $D_t^{\mathrm{fix}}$: baseline level of non-price-responsive demand in time block $t$ (MW)
\item[] $D^{\mathrm{res}}$: amount of price-responsive demand (MW)
\item[] $D_{s}^+$: positive demand shift under demand shifter $s$ (MW)
\item[] $D_{f}^-$: negative demand shift under fuel price shifter $f$ (MW)
\item[] $A_{grt}$: availability of generator $g$ in time block $t$ under profile $r$ (\%)
\item[] $\alpha_a$: tail probability at which CVaR is evaluated by market participant $a$, $0< \alpha_a \leq 1$
\item[] $\beta_a$: weight given to expected value of surplus in risk measure for market participant $a$, $0\leq \beta_a \leq 1$ 
 \item[] $p_{frs}$: nominal probability of scenario $(f,r,s)$ 
\item[] $\underline{v}_a$ / $ \overline{v}_a$: minimum/maximum volume of contract to be traded by market participant $a$ 
\end{itemize} \bigskip
\noindent\textit{Provisional parameters (i.e., values calculated by agents):}
\begin{itemize}
\item[] $\lambda_{frst}$: price of energy in time block $t$ under scenarios $(f,r,s)$ (\$/MWh)  
\item[] $\pi_{gfrst}$: profit per available unit of generation $g$ under scenarios $(f,r,s)$ in time block $t$ (\$/MW)  
\item[] $\phi^k$: price of contract $k$ incurred in the first stage (\$/MW-yr) 
\item[] $\eta_{frs}^k$: payout of contract $k$ under scenario $(f,r,s)$ (\$/MW-yr) 
\end{itemize} \bigskip
\noindent\textit{Variables:}
\begin{itemize}
\item[] $x_g$: capacity installed for generation technology $g$ (MW)
\item[] $y_{gt}$: power production by generation $g$ in time block $t$ (MW)
\item[] $d_{frst}^{\mathrm{fix}}$: amount of served non-price-responsive demand in time block $t$ under scenario $(f,r,s)$ (MW)
\item[] $d_{frst}^{\mathrm{res}}$: amount of served price-responsive demand in time block $t$ under scenario $(f,r,s)$ (MW)
\item[] $\rho_a$: risk measure for market participant $a$ (\$)
\item[] $v^k_a$: volume of contract $k$ sold or purchased by market participator $a$ (MW) 
\item[] VaR$_a$: value at risk for market participator $a$ (\$) 
\item[] $u_{frs}^{a}$: surplus for market participator $a$ under scenarios $(f,r,s)$ (\$) 
\item[] $u_{frs}^{a+}$: auxiliary variable of market participator $a$ for calculating VaR (\$)
\end{itemize}
\doublespacing

\subsection{Economic Dispatch}
The economic dispatch (ED) problem for each scenario is to find the power output of all generators leading to the lowest possible operating cost while maintaining system constraints. It is modeled as a convex quadratic problem. The outputs of the ED problems are parameters that enter the generator model and consumer model.
\begin{subequations} \label{eq:ED}
\begin{alignat}{2}
\text{(ED)}_{frs} & \quad H_{frs} =  \underset{y, d}{\operatorname{maximize}} & \quad \sum_{t \in \mathcal{T}} L_{t} B\left(d_{frst}^{\mathrm{fix}}+d_{frst}^{\mathrm{res}}-\left(d_{frst}^{\mathrm{res}}\right)^{2} /\left(2 D^{\mathrm{res}}\right)\right) \notag \\
& \quad &  -\sum_{t \in \mathcal{T}} \sum_{g \in G} L_{t} C_{gf}^{\mathrm{EN}} y_{gfrst} \label{eq:ED_obj}\\ 
&\text{subject to} \quad \quad (\lambda_{frst}): & d_{frst}^{\mathrm{fix}}+d_{frst}^{\mathrm{res}} + D^{+}_{s} - D^{-}_{f}=\sum_{g \in G} y_{gfrst} &\qquad \forall t \in \mathcal{T} \label{eq:ED_b}\\
& \quad \quad \quad \quad\quad \quad (\pi_{gfrst}): &   0 \leq y_{gfrst} \leq A_{grt} x_{g} &\qquad \forall g \in G, t \in \mathcal{T}  \label{eq:ED_c}\\
& \quad& 0 \leq d_{frst}^{\mathrm{fix}} \leq D_{t}^{\mathrm{fix}} &\qquad \forall t \in \mathcal{T} \label{eq:ED_d}\\ 
& \quad &  0 \leq d_{frst}^{\mathrm{res}} \leq D^{\mathrm{res}} &\qquad \forall t \in \mathcal{T}.  \label{eq:ED_e}
\end{alignat}
\end{subequations}

The objective function \eqref{eq:ED_obj} maximizes the total social surplus, which is equal to consumer surplus minus the total cost of production. Served demand is split into $d_{frst}^{\mathrm{fix}}, d_{frst}^{\mathrm{res}}, D^{+}_{s}, D^{-}_{f}$. As discussed in~\citet{mays2019asymmetric}, we treat $D^{+}_{s}, D^{-}_{f}$ as exogenous shocks to demand and exclude them from Eq.~\eqref{eq:ED_obj}, implying that serving it does not bring gain or loss to customers. Constraint \eqref{eq:ED_b} maintains power balance, i.e., the total demand served equals total generation. Constraint \eqref{eq:ED_c} states that power production is limited by the installed capacity of a given technology multiplied by its availability in the time segment. Constraints \eqref{eq:ED_d} and \eqref{eq:ED_e} model non-price-responsive and price-responsive demand. Note that $\lambda_{frst}$ is the dual of \eqref{eq:ED_b} and gives the energy price in the spot market, and $\pi_{gfrst}$ is the dual of \eqref{eq:ED_c}, representing the operating profit of generator $g$ per available unit of capacity.  

\subsection{Contracts}
Market participants can trade a number of instruments to hedge risk. Here we provide a more detailed description of the four instruments implemented in the numerical examples. The first three instruments are tailed to the risk profiles of each generation technology. We denote by $\eta^k_{frs}$ the payout of contract $k$ under scenario $(f,r,s)$.

Call options give the holder the right but not the obligation to obtain the specified energy at a specified strike price. When $k$ indexes a call option with strike price $\lambda^k$ that covers all the time blocks in a year, the payout is calculated as
\begin{align}
\eta^k_{frs} = \sum_{t \in \mathcal{T}} L_t \max\{ 0, \lambda_{frst} - \lambda^k\}.
\end{align}

When $k$ indexes a futures contract that covers all the time blocks at price $\lambda^k$, the payout is calculated as
\begin{align}
\eta^k_{frs} = \sum_{t \in \mathcal{T}} L_t ( \lambda_{frst} - \lambda^k).
\end{align}

A unit contingent contract is defined to track the availability of variable technologies so that the payout is proportional to the output. When $k$ indexes a unit contingent contract at price $\lambda^k$ for generator $g$, the payout is calculated as
\begin{align}
\eta^k_{frs} = \sum_{t \in \mathcal{T}}A_{grt} L_t (\lambda_{frst} - \lambda^k).
\end{align}

The SFPFC is defined to track a shape determined ex post by the demand for energy. When $k$ indexes an SFPFC that covers a whole year at price $\lambda^k$, the payout is calculated as
\begin{align}
\eta^k_{frs} =  \sum_{t \in \mathcal{T}}  \frac{8760 (D_t^{\mathrm{fix}}+D^+_s - D^-_f)}{\sum_{t' \in \mathcal{T}}L_{t'}(D_{t'}^{\mathrm{fix}}+D^+_s - D^-_f)} L_t ( \lambda_{frst} - \lambda^k).
\end{align}
We note that our definition of the SFPFC shape excludes price-responsive load, but includes all fixed load regardless of whether it is served.

\subsection{Generator Model}
A single representative agent is used to model investing in each generation technology. Generators make capacity decisions and sell contracts in the first stage and produce energy in the second stage. The generator model is stated as follows:
\begin{subequations} \label{eq:GEN}
\begin{alignat}{2}
\text{(GEN)}_g \quad \rho_g = & & & \nonumber \\
\underset{v_{g}, u^{g}, u^{g+}, \mathrm{VaR}_{g}}{\operatorname{maximize}} & \quad & \left(1-\beta_{g}\right)\left(\mathrm{VaR}_{g}-1 / \alpha_{g} \sum_{f \in \mathcal{F}} \sum_{r \in \mathcal{R}} \sum_{s \in S} p_{frs} u^{g+}_{frs}\right) \notag \\ & \quad &+\beta_{g} \sum_{f \in \mathcal{F}} \sum_{r \in \mathcal{R}} \sum_{s \in \mathcal{S}} p_{frs}u^{g}_{frs} - \frac{ \gamma}{2} \sum_{k \in \mathcal{K}} (\sum_{a \in A} v_a^k)^2 \label{eq:GEN_obj}\\ 
\text {subject to} & \quad &  u^{g}_{frs}= \sum_{t \in \mathcal{T}} \pi_{gfrst} A_{grt} x_g -C_{g}^{\mathrm{INV}} x_{g} - \sum_{k \in K} v_g(\phi^k - \eta_{frs}^k) & \qquad \forall f \in \mathcal{F}, r \in \mathcal{R}, s \in \mathcal{S} \label{eq:GEN_b}\\
(\tau^g_{frs}): &  \quad &  \mathrm{VaR}_{g}-u^{g}_{frs} \leq u^{g+}_{frs} & \qquad \forall f \in \mathcal{F}, r \in \mathcal{R}, s \in \mathcal{S} \label{eq:GEN_c}\\
 & \quad & 0 \leq u^{g+}_{frs} & \qquad \forall f \in \mathcal{F}, r \in \mathcal{R}, s \in \mathcal{S} \label{eq:GEN_d}\\
 & \quad & \underline{v}^k_g \leq v^k_g  \leq \overline{v}^k_g & \qquad \forall k \in \mathcal{K}. \label{eq:GEN_e}
\end{alignat}
\end{subequations}
The objective function \eqref{eq:GEN_obj} is maximizing the risk-adjusted profits for each generation technology $g$. The first two terms in \eqref{eq:GEN_obj} are the sum of a convex combination of expected value and CVaR of surplus, while the third term is penalizes imbalances in contract volumes and is zero in equilibrium. Constraint~\eqref{eq:GEN_b} states that the overall utility of each scenario is equal to operating profits in the scenario minus the investment cost minus the result of financial trades. Constraints~\eqref{eq:GEN_c} and~\eqref{eq:GEN_d} dictate the auxiliary variables $u_{frs}^{g+}$ used in calculating CVaR. Constraint \eqref{eq:GEN_e} limits the quantity of financial trading.

\subsection{Consumer Model}
Here a single representative consumer is modeled, with its decisions equivalent to a number of identical small consumers in a perfectly competitive market. The consumer signs contacts with generators in the first stage and consumes energy in the second stage. The consumer model is stated as follows:
\begin{subequations}
\begin{alignat}{2}
\text{(CON)} \quad \rho_c = &&& \nonumber \\
\underset{u^{c}, u^{c+}, \mathrm{VaR}_{c}}{\operatorname{maximize}} & \quad & \left(1-\beta_{c}\right)\left(\mathrm{VaR}_{c}-1 / \alpha_{c} \sum_{f \in \mathcal{F}} \sum_{r \in \mathcal{R}} \sum_{s \in \mathcal{S}} p_{frs} u^{c+}_{frs}\right) \notag \\ 
& & +\beta_{c} \sum_{f \in \mathcal{F}} \sum_{r \in \mathcal{R}} \sum_{s \in \mathcal{S}} p_{frs} u^{c}_{frs} \label{eq:CON_obj}\\
\text {subject to} \: &\quad & u^{c}_{frs}= \sum_{t \in \mathcal{T}} L_{t} B\left(d_{frst}^{\mathrm{fix}}+d_{frst}^{\mathrm{res}}-\left(d_{frst}^{\mathrm{res}}\right)^{2} /\left(2 D^{\mathrm{res}}\right)\right) \nonumber \\
& \quad & -\sum_{t \in \mathcal{T}} L_{t} \lambda_{frst} (d_{frst}^{\mathrm{fix}}+d_{frst}^{\mathrm{res}} + D^{+}_{s}-D^{-}_{f}) \nonumber\\
&\quad& -\sum_{k \in K}v_c(\phi^k-\eta^k_{frs}) & \qquad  \forall f \in \mathcal{F}, r \in \mathcal{R}, s \in \mathcal{S} \label{eq:CON_b}\\ 
& \quad & \mathrm{VaR}_{c}-u^{c}_{frs} \leq u^{c+}_{frs} & \qquad \forall f \in \mathcal{F}, r \in \mathcal{R}, s \in \mathcal{S}  \label{eq:CON_c}\\  
& \quad & 0 \leq u^{c+}_{frs} & \qquad \forall  f \in \mathcal{F}, r \in \mathcal{R}, s \in \mathcal{S} \label{eq:CON_d}\\
& \quad & \underline{v}^k_c \leq v^k_c  \leq \overline{v}^k_c & \qquad \forall k \in \mathcal{K}. \label{eq:CON_e}
\end{alignat}
\end{subequations}

The objective function \eqref{eq:CON_obj} maximizes the risk-adjusted utility for consumers, i.e., the sum of a convex combination of expected value and CVaR of surplus. Here, we do not include the third contract balance term as in~\eqref{eq:GEN_obj}, which equals zero in equilibrium and thus does not affect interpretation. Omission of this term led to faster convergence in numerical experiments. Constraint \eqref{eq:CON_b} states that the overall utility of each scenario is equal to the value of energy consumption minus the result of financial trades in scenario minus payments for energy. Constraint~\eqref{eq:CON_c} and~\eqref{eq:CON_d} dictate the auxiliary variables $u_{frs}^{c+}$ used in calculating CVaR. Constraint \eqref{eq:CON_e} limits the quantity of financial trading.

\subsection{Reliability credit}
In the mandatory trading example, a reliability credit is introduced to calculate the maximum quantity of options that generators can sell. Here we calculate the reliability credit as the resource's expected availability when all generation is deployed to its maximum availability. This definition can be interpreted as the marginal contribution that adding one unit of generation would make to the system's effective load carrying capability.
\begin{align}
    \text{(Reliability credit)} \quad rc_g = \frac{\sum_{f \in \mathcal{F}}\sum_{r \in \mathcal{R}}\sum_{s \in \mathcal{S}} L_t A_{grt} \mathbbm{1}\{\lambda_{frst} > C_{\text{`peaking'}f}^{\mathrm{EN}}\}}{\sum_{f \in \mathcal{F}}\sum_{r \in \mathcal{R}}\sum_{s \in \mathcal{S}} L_t\mathbbm{1}\{\lambda_{frst} > C_{\text{`peaking'}f}^{\mathrm{EN}}\}},
\end{align}
where $\mathbbm{1}\{\cdot \}$ is an indicator function.

\subsection{Algorithm}
The multi-agent equilibrium problem requires simultaneous solution of the economic dispatch, generator model, and consumer model. The equilibrium conditions are as follows:
\begin{subequations}
\begin{align}
    \sum_{a \in A} v_{a}^{k}=0 \quad \forall k \in \mathcal{K}; \label{eq:equ_a}\\
    0 \leq x_{g} \perp \rho_g \geq 0 \quad \forall g \in \mathcal{G}. \label{eq:equ_b}
\end{align}
\end{subequations}
Equation \eqref{eq:equ_a} describes that every financial trade must be balanced, i.e., the amount of contracts sold by the generator should equal to the contract bought by the consumer. Complementarity condition \eqref{eq:equ_b} is obtained from KKT conditions and can be interpreted as investment in technology $g$ will continue until it no longer results in risk-adjusted profit (\cite{ehrenmann2011generation}). The two conditions are used as the inner-loop and outer-loop stopping criteria respectively in the algorithm. 

Here we implement an ADMM-style algorithm to solve this problem. Note that $\varepsilon_1$ and $\varepsilon_2$ are step sizes that used to update capacity and contract price, while $\sigma$ and $\delta$ are small constants that are used as stopping criteria.
\begin{algorithm} 
	\caption{} 
	\label{alg:eq} 
	\begin{algorithmic} 
		\REQUIRE An instance of equilibrium problem $(EQ)$ defined by models~$(ED)$, $(CON)$, and $(GEN)$. 
		\ENSURE near-equilibrium solution to $(EQ)$
		\STATE{define $\sigma, \delta, \varepsilon_1, \varepsilon_2 > 0$; let $\rho_a = 0 \: \forall a \in \mathcal{A}$; initialize $x, \phi$}
		\LOOP 
		\STATE$x_g \leftarrow \max\{0,x_g + \varepsilon_1 \rho_g /C_g^{INV} \} \quad \forall g \in \mathcal{G}$ 
    \STATE{solve $(ED)_{frs}$; update $\lambda_{frst}, \pi_{gfrst},\eta_{frs}^k \: \forall (f,r,s) \in \mathcal{F}\times \mathcal{R}\times \mathcal{S}, k \in \mathcal{K}$}
      \STATE{solve $\text{(Reliability credit)}$; update $rc_g$ (only for mandatory case)}
      \STATE{Set the trading volume constraints}
		
		\STATE{solve $(CON)$}
		\STATE{solve $(GEN)_g \quad \forall g \in \mathcal{G}$}
		\WHILE{$\max_{k \in \mathcal{K}} \lvert \sum_{a \in \mathcal{A}} v_a^k \rvert > \sigma$}
		\STATE{$\phi^k \leftarrow \phi^k + \varepsilon_2 \sum_{a \in \mathcal{A}} v_a^k \quad \forall k \in \mathcal{K}$}
		\STATE{solve $(CON)$; update $v_c^k$}
		\STATE{solve $(GEN)_g \quad \forall g \in \mathcal{G}$; update $v_g^k$}
		\ENDWHILE
		\IF{$\max_{g \in \mathcal{G}} \lvert \rho_g \rvert < \delta$}
		\STATE return $x$ and $\phi$
		\ENDIF
		\ENDLOOP
	\end{algorithmic}
\end{algorithm}

For generators selling SFPFC collectively, some modifications are made in the generator model and the algorithm. A single agent $N$ is employed to represent all generation companies and trade with consumers. The generator model is in the same form as \eqref{eq:GEN} except that its overall utility of each scenario is the sum of surplus of all generation technologies minus the result of financial trades as a whole.
\begin{subequations}
\begin{alignat}{2}
\text{(GEN)}_N \quad \rho_N = & & & \nonumber \\
\underset{v_{N}, u^{N}, u^{N+}, \mathrm{VaR}_{N}}{\operatorname{maximize}} & \quad & \left(1-\beta_{N}\right)\left(\mathrm{VaR}_{N}-1 / \alpha_{N} \sum_{f \in \mathcal{F}} \sum_{r \in \mathcal{R}} \sum_{s \in S} p_{frs} u^{N+}_{frs}\right) \notag \\ & \quad &+\beta_{N} \sum_{f \in \mathcal{F}} \sum_{r \in \mathcal{R}} \sum_{s \in \mathcal{S}} p_{frs}u^{N}_{frs} - \frac{ \gamma}{2} (\sum_{a \in A} v_a)^2 \\ 
\text {subject to} &\quad & u^{N}_{frs}= \sum_{g \in \mathcal{G}}\sum_{t \in \mathcal{T}} \pi_{gfrst} A_{grt} x_g -\sum_{g \in \mathcal{G}}C_{g}^{\mathrm{INV}} x_{g} \nonumber \\
&\quad &  - v_N(\phi - \eta_{frs})   & \qquad \forall f \in \mathcal{F}, r \in \mathcal{R}, s \in \mathcal{S} \\
(\tau^N_{frs}): &\quad  &  \mathrm{VaR}_{N}-u^{N}_{frs} \leq u^{N+}_{frs} & \qquad \forall f \in \mathcal{F}, r \in \mathcal{R}, s \in \mathcal{S} \label{eq:COLL_c}\\
&\quad & 0 \leq u^{N+}_{frs} & \qquad \forall f \in \mathcal{F}, r \in \mathcal{R}, s \in \mathcal{S} \\
&\quad & \underline{v}_N \leq v_N  \leq \overline{v}_N.
\end{alignat}
\end{subequations}

When generators sell SFPFC collectively, the step in the algorithm for updating capacity is replaced with
\begin{align}
    &x_g \leftarrow \max\{0,x_g + \varepsilon_1 \psi_g /C_g^{INV} \} &\quad \forall g \in \mathcal{G}, 
\end{align}
where $ \psi_g = \sum_{f \in \mathcal{F}}\sum_{ r \in \mathcal{R}}\sum_{s \in \mathcal{S}} (\tau^N_{frs}+\beta_N \cdot p_{frs})(-C_{g}^{\mathrm{INV}} x_{g} + \sum_{t \in \mathcal{T}} \pi_{gfrst} A_{grt} x_g )$. Note that $\tau^N_{frs}$ is the dual variable of constraint \eqref{eq:COLL_c} and $\tau^N_{frs}+\beta_N \cdot p_{frs}$ implies placing higher weights to worst-case scenarios where~$\tau^N_{frs}$ is nonzero. Hence $\psi_g$ represents the risk-adjusted profit of technology $g$ without any effect of financial trading.






\newpage

\bibliography{ref.bib}

\begin{thebibliography}{}

\bibitem[\protect\citeauthoryear{Abada, de~Maere~d'Aertrycke, and Smeers}{Abada
  et~al.}{2017}]{Abada2017}
Abada, I., G.~de~Maere~d'Aertrycke, and Y.~Smeers (2017).
\newblock On the multiplicity of solutions in generation capacity investment
  models with incomplete markets: a risk-averse stochastic equilibrium
  approach.
\newblock {\em Mathematical Programming\/}~{\em 165\/}(1), 5--69.

\bibitem[\protect\citeauthoryear{{AEMO}}{{AEMO}}{2022}]{AEMO}
{AEMO} (2022, June).
\newblock {AEMO} suspends {NEM} wholesale market.
\newblock Available at
  \url{https://aemo.com.au/newsroom/media-release/aemo-suspends-nem-wholesale-market}.

\bibitem[\protect\citeauthoryear{Artzner, Delbaen, Eber, and Heath}{Artzner
  et~al.}{1999}]{Artzner1999}
Artzner, P., F.~Delbaen, J.~Eber, and D.~Heath (1999).
\newblock Coherent measures of risk.
\newblock {\em Mathematical Finance\/}~{\em 9\/}(3), 203--228.

\bibitem[\protect\citeauthoryear{Baker and Coleman}{Baker and
  Coleman}{2022}]{baker2022paying}
Baker, C.~M. and J.~W. Coleman (2022).
\newblock Paying for energy peaks: Learning from {Texas}' {February} 2021 power
  crisis.
\newblock {\em Transactions: The Tennessee Journal of Business Law\/}~{\em
  23\/}(3), 7.

\bibitem[\protect\citeauthoryear{Batlle, Schittekatte, and Knittel}{Batlle
  et~al.}{2022a}]{batlle2022b}
Batlle, C., T.~Schittekatte, and C.~R. Knittel (2022a).
\newblock Power price crisis in the {EU} 2.0+: Desperate times call for
  desperate measures.
\newblock Available at SSRN: \url{https://dx.doi.org/10.2139/ssrn.4074014}.

\bibitem[\protect\citeauthoryear{Batlle, Schittekatte, and Knittel}{Batlle
  et~al.}{2022b}]{batlle2022a}
Batlle, C., T.~Schittekatte, and C.~R. Knittel (2022b).
\newblock Unveiling current policy responses and proposing a balanced
  regulatory remedy.
\newblock Available at
  \url{https://energy.mit.edu/wp-content/uploads/2022/02/MITEI-WP-2022-02.pdf}.

\bibitem[\protect\citeauthoryear{Bezanson, Edelman, Karpinski, and
  Shah}{Bezanson et~al.}{2017}]{Bezanson2017Julia}
Bezanson, J., A.~Edelman, S.~Karpinski, and V.~B. Shah (2017).
\newblock Julia: A fresh approach to numerical computing.
\newblock {\em SIAM Review\/}~{\em 59\/}(1), 65--98.

\bibitem[\protect\citeauthoryear{Billimoria, Fele, Savelli, Morstyn, and
  McCulloch}{Billimoria et~al.}{2022}]{Billimoria2022}
Billimoria, F., F.~Fele, I.~Savelli, T.~Morstyn, and M.~McCulloch (2022).
\newblock An insurance mechanism for electricity reliability differentiation
  under deep decarbonization.
\newblock {\em Applied Energy\/}~{\em 321}, 119356.

\bibitem[\protect\citeauthoryear{Billimoria and Poudineh}{Billimoria and
  Poudineh}{2019}]{Billimoria2019}
Billimoria, F. and R.~Poudineh (2019).
\newblock Market design for resource adequacy: A reliability insurance overlay
  on energy-only electricity markets.
\newblock {\em Utilities Policy\/}~{\em 60}, 100935.

\bibitem[\protect\citeauthoryear{Bothwell and Hobbs}{Bothwell and
  Hobbs}{2017}]{bothwell2017crediting}
Bothwell, C. and B.~F. Hobbs (2017).
\newblock Crediting wind and solar renewables in electricity capacity markets:
  the effects of alternative definitions upon market efficiency.
\newblock {\em The Energy Journal\/}~{\em 38\/}(KAPSARC Special Issue).

\bibitem[\protect\citeauthoryear{Cramton, Ockenfels, and Stoft}{Cramton
  et~al.}{2013}]{Cramton2013}
Cramton, P., A.~Ockenfels, and S.~Stoft (2013).
\newblock Capacity market fundamentals.
\newblock {\em Economics of Energy {\&} Environmental Policy\/}~{\em 2\/}(2),
  27--46.

\bibitem[\protect\citeauthoryear{de~Maere~d’Aertrycke, Ehrenmann, Ralph, and
  Smeers}{de~Maere~d’Aertrycke et~al.}{2017}]{DAertrycke2017wp}
de~Maere~d’Aertrycke, G., A.~Ehrenmann, D.~Ralph, and Y.~Smeers (2017,
  December).
\newblock Risk trading in capacity equilibrium models.
\newblock Available at \url{https://doi.org/10.17863/CAM.17552}.

\bibitem[\protect\citeauthoryear{Dunning, Huchette, and Lubin}{Dunning
  et~al.}{2017}]{DunningHuchetteLubin2017JuMP}
Dunning, I., J.~Huchette, and M.~Lubin (2017).
\newblock Jump: A modeling language for mathematical optimization.
\newblock {\em SIAM Review\/}~{\em 59\/}(2), 295--320.

\bibitem[\protect\citeauthoryear{Ehrenmann and Smeers}{Ehrenmann and
  Smeers}{2011}]{ehrenmann2011generation}
Ehrenmann, A. and Y.~Smeers (2011).
\newblock Generation capacity expansion in a risky environment: a stochastic
  equilibrium analysis.
\newblock {\em Operations research\/}~{\em 59\/}(6), 1332--1346.

\bibitem[\protect\citeauthoryear{{European Commission}}{{European
  Commission}}{2022}]{EUR_COMMISSION}
{European Commission} (2022, May).
\newblock Short-term energy market interventions and long term improvements to
  the electricity market design – a course for action.
\newblock Available at
  \url{https://www.astrid-online.it/static/upload/com_/com_2022_236_1_en_act_part1_v7.pdf}.

\bibitem[\protect\citeauthoryear{Ferris and Philpott}{Ferris and
  Philpott}{2022}]{Ferris2022}
Ferris, M. and A.~Philpott (2022).
\newblock Dynamic risked equilibrium.
\newblock {\em Operations Research\/}~{\em 70\/}(3), 1933--1952.

\bibitem[\protect\citeauthoryear{G{\'e}rard, Lecl{\`e}re, and
  Philpott}{G{\'e}rard et~al.}{2018}]{Gerard2018}
G{\'e}rard, H., V.~Lecl{\`e}re, and A.~Philpott (2018).
\newblock On risk averse competitive equilibrium.
\newblock {\em Operations Research Letters\/}~{\em 46\/}(1), 19--26.

\bibitem[\protect\citeauthoryear{Gruber, Gauster, Laaha, Regner, and
  Schmidt}{Gruber et~al.}{2022}]{Gruber2022}
Gruber, K., T.~Gauster, G.~Laaha, P.~Regner, and J.~Schmidt (2022, May).
\newblock Profitability and investment risk of texan power system
  winterization.
\newblock {\em Nature Energy\/}~{\em 7\/}(5), 409--416.

\bibitem[\protect\citeauthoryear{{Gurobi Optimization, LLC}}{{Gurobi
  Optimization, LLC}}{2020}]{gurobi}
{Gurobi Optimization, LLC} (2020).
\newblock Gurobi optimizer reference manual.

\bibitem[\protect\citeauthoryear{Hogan}{Hogan}{2005}]{hogan2005energy}
Hogan, W.~W. (2005).
\newblock On an “energy only” electricity market design for resource
  adequacy.
\newblock Available at
  \url{https://scholar.harvard.edu/whogan/files/hogan_energy_only_092305.pdf}.

\bibitem[\protect\citeauthoryear{Joskow}{Joskow}{2008}]{joskow2008capacity}
Joskow, P.~L. (2008).
\newblock Capacity payments in imperfect electricity markets: Need and design.
\newblock {\em Utilities Policy\/}~{\em 16\/}(3), 159--170.

\bibitem[\protect\citeauthoryear{Joskow}{Joskow}{2022}]{Joskow2022}
Joskow, P.~L. (2022).
\newblock From hierarchies to markets and partially back again in electricity:
  responding to decarbonization and security of supply goals.
\newblock {\em Journal of Institutional Economics\/}~{\em 18\/}(2), 313–329.

\bibitem[\protect\citeauthoryear{Macey and Ward}{Macey and
  Ward}{2021}]{Macey2021}
Macey, J.~C. and R.~Ward (2021).
\newblock {MOPR} madness.
\newblock {\em Energy Law Journal\/}~{\em 42\/}(1), 67--122.

\bibitem[\protect\citeauthoryear{Mays, Craig, Kiesling, Macey, Shaffer, and
  Shu}{Mays et~al.}{2022}]{mays2022private}
Mays, J., M.~T. Craig, L.~Kiesling, J.~C. Macey, B.~Shaffer, and H.~Shu (2022).
\newblock Private risk and social resilience in liberalized electricity
  markets.
\newblock {\em Joule\/}~{\em 6\/}(2), 369--380.

\bibitem[\protect\citeauthoryear{Mays, Morton, and O’Neill}{Mays
  et~al.}{2019}]{mays2019asymmetric}
Mays, J., D.~P. Morton, and R.~P. O’Neill (2019).
\newblock Asymmetric risk and fuel neutrality in electricity capacity markets.
\newblock {\em Nature Energy\/}~{\em 4\/}(11), 948--956.

\bibitem[\protect\citeauthoryear{Moreno, Barroso, Rudnick, Mocarquer, and
  Bezerra}{Moreno et~al.}{2010}]{Moreno2010}
Moreno, R., L.~Barroso, H.~Rudnick, S.~Mocarquer, and B.~Bezerra (2010).
\newblock Auction approaches of long-term contracts to ensure generation
  investment in electricity markets: Lessons from the brazilian and chilean
  experiences.
\newblock {\em Energy Policy\/}~{\em 38\/}(10), 5758--5769.

\bibitem[\protect\citeauthoryear{Muñoz, Suazo-Martínez, Pereira, and
  Moreno}{Muñoz et~al.}{2021}]{Munoz2021}
Muñoz, F.~D., C.~Suazo-Martínez, E.~Pereira, and R.~Moreno (2021).
\newblock Electricity market design for low-carbon and flexible systems: Room
  for improvement in chile.
\newblock {\em Energy Policy\/}~{\em 148}, 111997.

\bibitem[\protect\citeauthoryear{Neuhoff and De~Vries}{Neuhoff and
  De~Vries}{2004}]{Neuhoff2004}
Neuhoff, K. and L.~De~Vries (2004).
\newblock Insufficient incentives for investment in electricity generations.
\newblock {\em Utilities Policy\/}~{\em 12\/}(4), 253--267.

\bibitem[\protect\citeauthoryear{Newbery}{Newbery}{2016}]{Newbery2016}
Newbery, D. (2016).
\newblock Missing money and missing markets: Reliability, capacity auctions and
  interconnectors.
\newblock {\em Energy Policy\/}~{\em 94}, 401--410.

\bibitem[\protect\citeauthoryear{Oren}{Oren}{2005}]{oren2005generation}
Oren, S.~S. (2005).
\newblock Generation adequacy via call options obligations: Safe passage to the
  promised land.
\newblock {\em The electricity journal\/}~{\em 18\/}(9), 28--42.

\bibitem[\protect\citeauthoryear{Philpott, Ferris, and Wets}{Philpott
  et~al.}{2016}]{Philpott2016}
Philpott, A., M.~Ferris, and R.~Wets (2016).
\newblock Equilibrium, uncertainty and risk in hydro-thermal electricity
  systems.
\newblock {\em Mathematical Programming\/}~{\em 157\/}(2), 483--513.

\bibitem[\protect\citeauthoryear{Ralph and Smeers}{Ralph and
  Smeers}{2015}]{Ralph2015}
Ralph, D. and Y.~Smeers (2015).
\newblock Risk trading and endogenous probabilities in investment equilibria.
\newblock {\em SIAM Journal on Optimization\/}~{\em 25}, 2589--2611.

\bibitem[\protect\citeauthoryear{Rockafellar and Uryasev}{Rockafellar and
  Uryasev}{2000}]{Rockafellar2000}
Rockafellar, R. and S.~Uryasev (2000).
\newblock Optimization of conditional value-at-risk.
\newblock {\em Journal of Risk\/}~{\em 2}, 21--41.

\bibitem[\protect\citeauthoryear{Schlag, Ming, Olson, Alagappan, Carron,
  Steinberger, and Jiang}{Schlag et~al.}{2020}]{schlag2020capacity}
Schlag, N., Z.~Ming, A.~Olson, L.~Alagappan, B.~Carron, K.~Steinberger, and
  H.~Jiang (2020).
\newblock Capacity and reliability planning in the era of decarbonization:
  Practical application of effective load carrying capability in resource
  adequacy.
\newblock {\em Energy and Environmental Economics, Inc.,. https://www. ethree.
  com/elcc-resource-adequacy\/}.

\bibitem[\protect\citeauthoryear{Simshauser}{Simshauser}{2019}]{simshauser2019stability}
Simshauser, P. (2019).
\newblock On the stability of energy-only markets with government-initiated
  contracts-for-differences.
\newblock {\em Energies\/}~{\em 12\/}(13), 2566.

\bibitem[\protect\citeauthoryear{Simshauser}{Simshauser}{2021}]{Simshauser2021}
Simshauser, P. (2021).
\newblock Vertical integration, peaking plant commitments and the role of
  credit quality in energy-only markets.
\newblock {\em Energy Economics\/}~{\em 104}, 105612.

\bibitem[\protect\citeauthoryear{Vazquez, Rivier, and Perez-Arriaga}{Vazquez
  et~al.}{2002}]{Vazquez2002}
Vazquez, C., M.~Rivier, and I.~Perez-Arriaga (2002).
\newblock A market approach to long-term security of supply.
\newblock {\em IEEE Transactions on Power Systems\/}~{\em 17\/}(2), 349--357.

\bibitem[\protect\citeauthoryear{Willems and Morbee}{Willems and
  Morbee}{2010}]{willems2010market}
Willems, B. and J.~Morbee (2010).
\newblock Market completeness: How options affect hedging and investments in
  the electricity sector.
\newblock {\em Energy Economics\/}~{\em 32\/}(4), 786--795.

\bibitem[\protect\citeauthoryear{Wolak}{Wolak}{2022}]{wolak2022long}
Wolak, F.~A. (2022).
\newblock Long-term resource adequacy in wholesale electricity markets with
  significant intermittent renewables.
\newblock {\em Environmental and Energy Policy and the Economy\/}~{\em 3\/}(1),
  155--220.

\end{thebibliography}

\newpage

\section*{Supplementary Information}

\subsection*{Supplementary Note 1: Complete Trading}
For comparison, we also construct a complete trading model, where the risk of market participants can be fully traded with no transaction cost. The complete trading equilibrium can be found through a risk-averse optimization using the intersection of risk sets of all agents in the model. Using the optimal objective $H_{frs}$ from model~$(ED)$, the complete trading model can be formulated as: 
\begin{subequations} \label{eq:SOC}
\begin{alignat}{2}
\text{(SOC)} &\quad  \rho_i = &&  \notag \\
& \underset{x, y, d, H, u^{i}, u^{i+}, \mathrm{VaR}_{i}}{\operatorname{maximize}} & \left(1-\beta_{i}\right)\left(\mathrm{VaR}_{i}-1 / \alpha_{i} \sum_{f \in \mathcal{F}} \sum_{r \in \mathcal{R}} \sum_{s \in \mathcal{S}} p_{frs} u_{frs}^{i+}\right) &+\beta_{i} \sum_{f \in \mathcal{F}} \sum_{r \in \mathcal{R}} \sum_{s \in \mathcal{S}} p_{frs} u_{frs}^{i}  \label{eq:SOC_obj}\\ 
& \quad \text { subject to } & u_{frs}^{i}=-\sum_{g \in \mathcal{G}} C_{g}^{\mathrm{INV}} x_{g}+H_{frs} & \qquad \forall f \in \mathcal{F}, r \in \mathcal{R}, s \in \mathcal{S} \label{eq:SOC_b}\\ 
& \quad& \mathrm{VaR}_{i}-u_{frs}^{i} \leq u_{frs}^{i+} & \qquad \forall f \in \mathcal{F}, r \in \mathcal{R}, s \in \mathcal{S} \label{eq:SOC_c}\\  
& \quad & 0 \leq u_{frs}^{i+} & \qquad \forall f \in \mathcal{F}, r \in \mathcal{R}, s \in \mathcal{S}. \label{eq:SOC_d}
\end{alignat}
\end{subequations}
The objective function \eqref{eq:SOC_obj} is maximizing the risk-adjusted net social surplus. Constraint \eqref{eq:SOC_b} states that the net social surplus of a scenario is equal to the total social surplus in operations minus the investment cost of generators. Constraint \eqref{eq:SOC_c} and \eqref{eq:SOC_d} dictate the auxiliary variables $u_{frs}^{i+}$ used in calculating CVaR. 

The risk parameters for the social optimization are determined by the intersection of the risk sets across all market participants. Since $\alpha$ is the same for all agents in our examples, the intersection is determined by the highest value of $\beta$ in the example; i.e., $\beta_i = \max \beta_a$. Thus, when $\beta=0.2,0.4$, or $0.6$ for generators, the retailer is less risk averse than the generators and we use the parameter $\beta=0.7$; when $\beta=0.8$ for the generators, they are less risk averse and we use the parameter $\beta=0.8$. The results of complete trading is shown in Table \ref{tab:complete}.
\begin{table}[b]
\captionsetup{justification=raggedright,singlelinecheck=false,labelfont=bf}
    \centering
    \caption{\textbf{Complete trading results.} Change in surplus is calculated against the three-contract unrestricted case at risk aversion level $\beta = 0.6$ and $\beta = 0.8$ respectively.}
    \begin{tabular}{c|cccc}
    \hline
        $\beta$ & Baseload(GW) & Peaker (GW) & Variable (GW) & Change of surplus (\$ M/yr) \\
        \hline
        $\beta = 0.7$ & 34.2 & 105.2 & 153.7 & 34\\
        $\beta = 0.8$ & 31.2 & 107.2 & 158.2 & 144 \\
        \hline
    \end{tabular}
    \label{tab:complete}
\end{table}

\subsection*{Supplementary Note 2: Multiple equilibria when selling SFPFC collectively}
We identified multiple equilibria when selling SFPFC collectively, with the equilibria identified by the algorithm depending on the starting point used. Change in surplus relative to the unrestricted case is consistent across the identified equilibria, leading to the same economic interpretation regardless of which is chosen. Nevertheless, characterizing these multiple equilibria represents an important topic for further research. The results presented in Table~\ref{tab:SFPFC} for selling SFPFC collectively are obtained using initial capacities equal to the unrestricted case at the same level of risk aversion. We repeat these results in Table~\ref{tab:SFPFC_col} as ``Initial~1.'' Other equilibria reported as ``Initial~2'' and ``Initial~3'' are obtained by setting starting points equal to the equilibrium capacities in the mandatory case and selling SFPFC separately case.

\begin{table}[]
\captionsetup{justification=raggedright,singlelinecheck=false,labelfont=bf}
    \centering
    \caption{\textbf{Multiple equilibria when selling SFPFC collectively.} Change in surplus is calculated against the unrestricted case and is roughly consistent across the identified equilibria for each level of risk aversion.} \label{tab:SFPFC_col}
    \begin{tabular}{l|cccccc}
\hline
\hline
&\multicolumn{3}{c}{$\beta = 0.2$} &\multicolumn{3}{c}{$\beta = 0.4$}\\
& Initial 1 & Initial 2 & Initial 3 & Initial 1 & Initial 2 & Initial 3  \\
\hline
Capacity (GW) \\
\quad Baseload &40.7 &51.9 &78.7 &34.6 &44.4 &49.8 \\
\quad Peaker   &99.9 &92.8 &77.2 &104.4 &97.7 &94.3 \\
\quad Variable &143.2 &119.4 &56.4 &153.6 &135.1 &123.7\\
\\
Trade Volume (GW) &91.6 &92.3 &93.6 &91.9 &92.3 &92.3\\
Contract risk premium (\$/MW-yr) & 19,039 &18,726 &17,029 &18,335 &18,176 &18,550 \\
\\
Average Price (\$/MWh)\\
\quad Spot &57.49 &57.53 &57.84 &57.39 &57.40 &57.38 \\
\quad Hedged &59.59 &59.61 &59.84 &59.43 &59.43 &59.44 \\
\\
 Interannual Volatility (\$/MWh)\\
\quad Spot &24.89 &24.44 &22.43 &24.43 &24.46 &24.22 \\
\quad Hedged &1.75 &1.62 &1.25 &1.69 &1.60 &1.61 \\
\\
Expected Unserved Energy (GWh) &5.83 &5.45 &4.20 &5.52 &5.51 &5.32 \\
Proximity to Equilibrium &0.016\% &0.017\% & 0.003\% &0.004\% & 0.017\% &0.012\% \\
Change in Surplus (\$M/yr) &-517 &-513 &-694 &-375 &-367 &-375\\
\hline
&\multicolumn{3}{c}{$\beta = 0.6$} &\multicolumn{3}{c}{$\beta = 0.8$}\\
& Initial 1 & Initial 2 & Initial 3 & Initial 1 & Initial 2 & Initial 3  \\
\hline
Capacity (GW) \\
\quad Baseload &33.6 &37.2 &54.9 &31.4 &31.5 &45.0\\
\quad Peaker   &105.4 &102.8 &91.3 &107.0 &106.9 &97.7\\
\quad Variable &154.3 &147.5 &111.2 &157.7 &157.4 &131.8\\
\\
Trade Volume (GW) &93.6  &93.6 &94.2 &95.8 &95.7 &96.2\\
Contract risk premium (\$/MW-yr) &15,866  &15,513 &14,800 &12,256 &12,231 &11742.2\\
\\
Average Price (\$/MWh)\\
\quad Spot &57.44 &57.48 &57.61 &57.59 &57.59 &57.68\\
\quad Hedged  &59.26 &59.25 &59.33 & 59.06 &59.06 &59.10\\
\\
Interannual Volatility (\$/MWh)\\
\quad Spot &24.20 &24.17 &24.11 &24.34 &24.34 &24.39\\
\quad Hedged &1.38 &1.39 &1.25 &1.14 &1.14 &1.10\\
\\
Unserved Energy (GWh) &5.35 &5.36 &5.21  &5.39 &5.39 &5.38\\
Proximity to Equilibrium &0.008\% &0.017\% &0.001\%  &0.012\% &0.017\% &0.017\%\\
Change in Surplus (\$M/yr) &-212 &-210 &-267 &-41 &-41 &-71\\
\hline
\hline
\end{tabular}
\end{table}

\subsection*{Supplementary Note 3: Options only}
To compare the benefits when options can be sold collectively, we list the results in Table \ref{tab:options}. Trading volumes when selling options collectively are always smaller than when selling options separately, but the gap is very small when $\beta=0.4$ and $\beta=0.6$. Allowing options to be sold collectively brings slight benefits in terms of surplus, expected unserved energy, average price and interannual volatility, but the benefits are not comparable with allowing selling SFPFC collectively.
\begin{table}[]
\captionsetup{justification=raggedright,singlelinecheck=false,labelfont=bf}
\centering
\caption{\textbf{Comparison of options when selling separately and collectively.} Change is surplus is calculated against the three-contract unrestricted case at the same level of risk aversion.} \label{tab:options}
\begin{tabular}{l|cccccc}
\hline
\hline
& \multicolumn{2}{c}{$\beta=0.2$} & \multicolumn{2}{c}{$\beta=0.4$} \\
& Sep & Col & Sep & Col  \\
\hline
Capacity (GW) \\
\quad Baseload &0.006 &9.7 &0.02 &3.1\\
\quad Peaker   &137.3 &130.0 &135.2 &133.0\\
\quad Variable &163.4 &149.3 &176.2 &171.0\\
\\
Trade Volume (GW)  &160.4 &158.5 &158.9 &158.6\\
Contract risk premium (\$/MW-yr) &2,671 &2,911 &2,643 &2,646 \\
\\
Average Price (\$/MWh)\\
\quad Spot &62.22 &62.15 &61.32 &61.27 \\
\quad Hedged &62.86 &62.82 &61.94 &61.90\\
\\
Interanuual Volatility (\$/MWh) \\
\quad Spot &25.86 &25.69 &25.42 &25.28 \\
\quad Hedged &13.79 &13.71 &13.63 &13.62 \\
\\
Expected Unserved Energy (GWh) &5.54 &5.44 &5.53 &5.46 \\
Proximity to Equilibrium &0.066\% &0.017\% &0.029\% &0.009\% \\
Change in Surplus (\$M/yr) &-5,042 &-4,998 &-4,206 &-4,165 \\
\hline
& \multicolumn{2}{c}{$\beta=0.6$} & \multicolumn{2}{c}{$\beta=0.8$}  \\
& Sep & Col & Sep & Col  \\
\hline
Capacity (GW) \\
\quad Baseload &0.2 &3.2 &12.4 &13.8\\
\quad Peaker    &133.6 &131.3 &122.6 &121.7\\
\quad Variable  &185.0 &180.4 &177.3  &174.9\\
\\
Trade Volume (GW)  &159.3 &159.2 &157.7  &156.5\\
Contract risk premium (\$/MW-yr) &2,650 &2,641 &2,473 &2,873 \\
\\
Average Price (\$/MWh)\\
\quad Spot &60.41 &60.36 &59.40 &59.31 \\
\quad Hedged  &61.03 &60.99 &59.98 &59.96 \\
\\
Interanuual Volatility (\$/MWh) \\
\quad Spot  &25.18 &25.11 &24.63 &24.47 \\
\quad Hedged  &13.61 &13.59 &13.45 &13.44 \\
\\
Expected Unserved Energy (GWh) &5.52 &5.47 &5.53 &5.44 \\
Proximity to Equilibrium &0.048\% &0.011\% &0.073\% &0.017\% \\
Change in Surplus (\$M/yr) &-3,411  &-3,367 &-2,448 &-2,431 \\
\hline
\hline
\end{tabular}
\end{table}

\subsection*{Supplementary Note 4: Unrestricted and mandatory SFPFC}
Besides allowing selling SFPFC collectively, here we also impose a mandatory requirement on purchase of SFPFCs, pushing the system to achieve higher reliability with moderate loss of surplus. The results are shown in Table~\ref{tab:SFPFC_2}. With a mandatory requirement, unserved energy is lower than both the three-contract case and the unrestricted SFPFC case but with lower surplus. As generators become less risk averse, the capacity mix becomes very close. 

\begin{table}
\captionsetup{justification=raggedright,singlelinecheck=false,labelfont=bf}
\begin{center}
\caption{\textbf{Effect of Mandatory Purchase of SFPFC.} Change in surplus is compared with the three-contract case.} \label{tab:SFPFC_2}
\begin{tabular}{l|ccccccc}
\hline 
\hline
& \multicolumn{2}{c}{$\beta=0.2$} & \multicolumn{2}{c}{$\beta=0.4$}  \\
 & Unrestricted & Mandatory & Unrestricted & Mandatory \\
\hline
Capacity (GW)   \\
\quad Baseload & 40.7  & 48.2 & 34.6 & 33.7 \\ 
\quad Peaker   & 99.9  & 96.2 & 104.4 & 105.8 \\ 
\quad Variable & 143.2  & 129.6 & 153.6 & 156.9 \\ 
Trade volume (GW) & 91.6  & 100.0 & 91.9 & 100.0 \\
Contract risk premium (\$/MW-yr) & 19039.4  & 51195.6 & 15866.1 & 42685.4\\
\\
Average Price (\$/MWh)\\
\quad Spot & 57.49  & 54.22 & 57.39 & 54.91 \\
\quad Hedged & 59.59  & 60.06 & 59.43 & 59.81 \\
\\
Interannual Volatility (\$/MWh)\\
\quad Spot & 24.89  & 20.48 & 24.43 & 21.26 \\
\quad Hedged & 1.75  & 1.11 & 1.69 & 1.18 \\
\\
Expected unserved energy (GWh) & 5.83  & 3.20 & 5.52 & 3.69 \\
Proximity to Equilibrium &  0.016\%  & 0.009\%  & 0.004\%  & 0.017\% \\
Change in Surplus (\$ M/yr) & -517  & -710 & -375& -540\\
\hline
 & \multicolumn{2}{c}{$\beta=0.6$} & \multicolumn{2}{c}{$\beta=0.8$}\\
 & Unrestricted & Mandatory & Unrestricted & Mandatory \\
\hline
Capacity (GW)   \\
\quad Baseload & 33.6  & 34.4 & 31.4 & 31.6 \\ 
\quad Peaker   & 105.4  & 105.2 & 107.0 & 107.0 \\ 
\quad Variable & 154.3  & 154.3 & 157.7 & 157.2 \\ 
Trade volume (GW) & 93.6  & 100.0 & 95.8 & 100.0 \\
Contract risk premium (\$/MW-yr)& 15866.1  & 29820.2 & 12256.2 & 15734.8\\
\\
Average Price (\$/MWh) \\
\quad Spot  & 57.44  & 56.01 & 57.59 & 57.23 \\
\quad Hedged & 59.26  & 59.49 & 59.06 &59.15 \\
\\
Interannual Volatility (\$/MWh) \\
\quad Spot  & 24.20  & 22.38 & 24.34 & 23.79 \\
\quad Hedged & 1.38  & 1.22 & 1.14 & 1.27 \\
\\
Expected unserved energy (GWh) & 5.35  & 4.30 & 5.39 & 5.06 \\
Proximity to Equilibrium &  0.008\%  & 0.009\%  & 0.012\%  & 0.024\% \\
Change in Surplus (\$ M/yr) & -212  & -302 & -41 & -70\\
\hline
\hline
\end{tabular}
\end{center}
\end{table}

\end{document}